\documentclass[aps,prl,twocolumn,amsmath,showpacs]{revtex4-1}
\usepackage{bm}
\usepackage{graphicx}
\begin{document}
\setlength{\arraycolsep}{2pt}

\title{Comment on "Role of Initial Entanglement and Non-Gaussianity in the Docoherence of Photon-Number Entangled States Evolving in a Noisy Channel"}
\author{Jaehak Lee$^{1}$, M. S. Kim$^{2}$, and Hyunchul Nha$^{1,3,*}$}
\affiliation{$^1$Department of Physics, Texas A \& M University at Qatar, Doha, Qatar\\
$^2$ QOLS, Blackett Laboratory,  Imperial College London, London SW7 2BW, United Kingdom\\
$^3$ School of Computational Sciences, Korea Institute for Advanced Study, Seoul, Korea}


\maketitle

In \cite{Allegra}, Allegra {\it et al.} employ several entanglement criteria, including the Simon criterion (SI), in order to provide evidence to support their conjecture that a Gaussian state remains entangled longer than a non-Gaussian state in a noisy Gaussian channel. In particular, they study the loss of entanglement for the class of photon number entangled states (PNES) $|\Psi\rangle=\sum_n\Psi_n|n\rangle|n\rangle$ in a thermal reservoir.
Here we show that their evidence is seriously flawed due to their use of entanglement criteria inappropriate for the comparison and that there exist a large class of non-Gaussian entangled states even within the PNES that can be more robust than Gaussian states.  

The dynamics of a system under two independent Markovian reservoirs can be described by
$\dot\rho=A\sum_{i=1,2} {\cal L}[a_i]\rho+B\sum_{i=1,2} {\cal L}[a_i^\dag]\rho$,
where $A$ and $B$ denote the interaction strength leading to dissipation and amplification, respectively (${\cal L}[O]\rho\equiv 2O\rho O^\dag-O^\dag O\rho-\rho O^\dag O$).
The case of $A>B$ describes the interaction with a thermal reservoir. Using the notations in [1], $A={\Gamma \over 2}(N_T+1)$ and $B={\Gamma \over 2}N_T$ ($N_T$: thermal photon number, $\Gamma$: decay rate).
Allegra {\it et al.} found that the Simon criterion (SI) based on the symplectic eigenvalues under partial transposition (PT) is optimal for PNES among the criteria they considered.  Here, in contrast to \cite{Allegra}, we employ the negativity of the density matrix under PT (NDPT) as the entanglement criterion and compare the entanglement dynamics of two particular PNES states studied in \cite{Allegra}: photon subtracted squeezed vacuum (PSSV : non-Gaussian) and the twin-beam state (TWB: Gaussian). For numerical purposes, we checked the negativity by restricting the elements of the whole density matrix to a subspace truncated by dimension $N_{\rm tr}$. It is well-known that the negativity under PT in a subspace is sufficient to verify entanglement. In Fig. 1 (a), we show the time to lose negativity for PSSV by NDPT ($N_{\rm tr}=3$) together with the separation time of TWB by SI (analytic result). These results clearly indicate that the PSSV remains entangled longer than the TWB for either the same initial entanglement $\epsilon_0$ or the same energy (indiscernible in this case).

The failure of Allegra {\it et al.} is attributed to the wrong choice of entanglement criteria; Markovian interaction would not create a new-type of correlation, thus, an initial entanglement witness will remain a significant tool to verify the entanglement of the decohered state at a later time. For a given state, it is usually nontrivial to identify the most efficient entanglement witness, however, it is obvious that the information on non-Gaussian entanglement is not fully contained in the covariance matrix (SI criterion). We instead checked the negativity of the decohered PNES under PT, which is a well-known tool particularly for continuous variables.
The NDPT can show the robustness of entanglement for a non-Gaussian PNES, even analytically. In \cite{Allegra}, they also studied randomly-generated PNES in a truncated basis, only to confirm their incorrect conclusion. Among these, the simplest one is $|\Phi\rangle_1=c_0|00\rangle+c_1|11\rangle$, for which the entanglement can be initially detected by the negativity in the subspace spanned by $|01\rangle$ and $|10\rangle$ under PT, which will be presumably useful also at later times. The separation time in this subspace can be analytically obtained by direct calculation. Fig. 1 (b) shows
the value of $B/A$ above which $|\Phi\rangle_1$ survives longer than the TWB with the same initial energy (blue) or entanglement (purple) as a function of $|c_1|^2$,
clearly demonstrating the failure of the evidence of Allegra {\it et al.} in a wide range of temperatures.
\begin{figure}[tbp]
\centerline{\scalebox{0.27}{\includegraphics[angle=270]{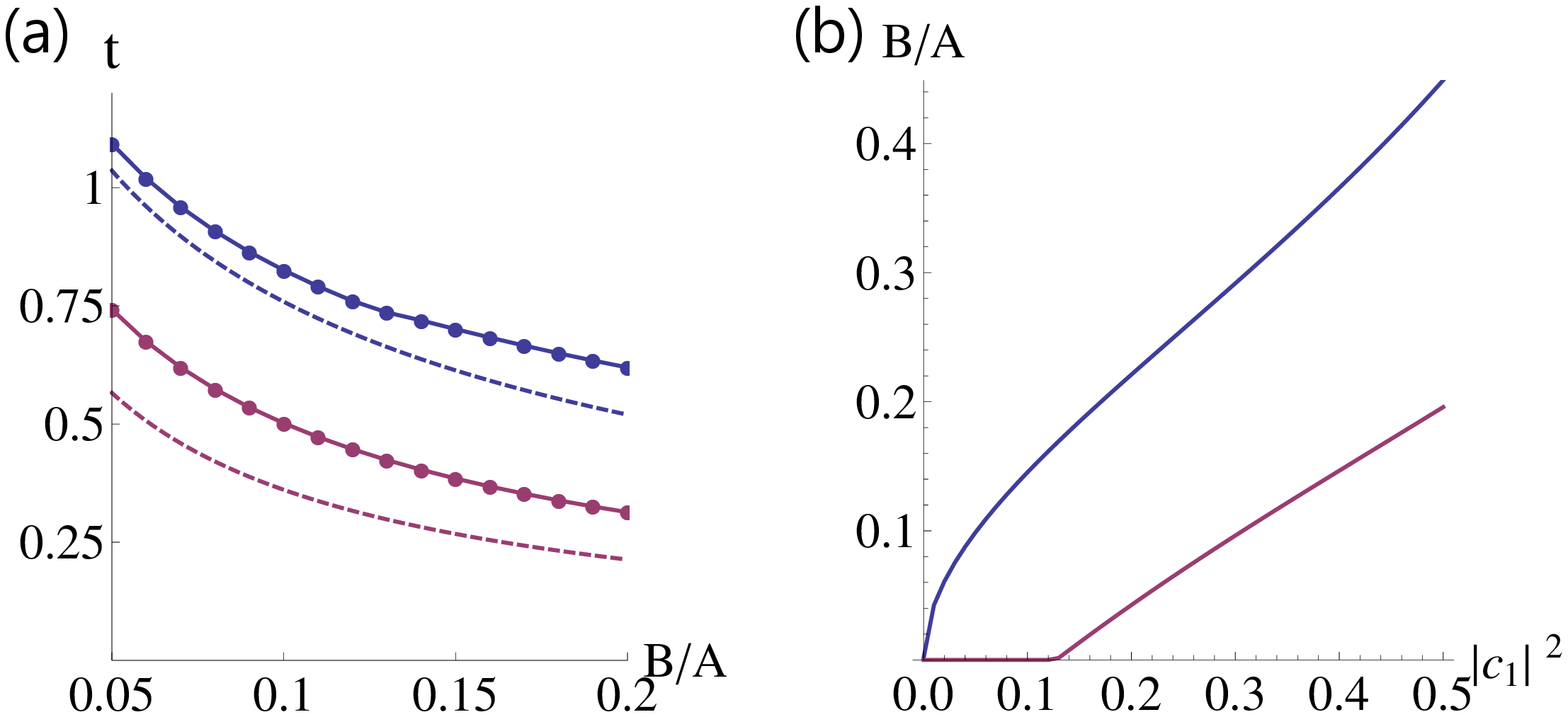}}}
\caption{(a) Time to lose the negativity in PT of PSSV (solid with circles) and the separation time of TWB (dotted) with $\epsilon_0=0.1$ (purple) and $\epsilon_0=1$ (blue) (b) the value of $B/A$ above which $c_0|00\rangle+c_1|11\rangle$ survives longer than TWB. }\end{figure}

In summary, we have shown that the evidence of Allegra {\it et al.} supporting their conjecture is derived from the wrong choice of entanglement criteria and disproved the maximal robustness of Gaussian entangled states.

Work supported by NPRP 4-554-1-084 from Qatar National Research Fund and UK EPSRC.
*hyunchul.nha@qatar.tamu.edu


\vspace{-0.5cm}\begin{thebibliography}{99}
\vspace{-0.5cm}
\bibitem{Allegra} M. Allegra, P. Giorda, and M. G. A. Paris, \prl  {\bf 105}, 100503 (2010).
\end{thebibliography}
\end{document}